\documentclass[aps,prl,twocolumn,nofootinbib,groupedaddress,amsfonts,floatfix]{revtex4}

\usepackage{graphicx,amsmath,amssymb,amstext}
\usepackage{amssymb,amsbsy,amsfonts,amsthm,hyperref}
\usepackage{lscape,placeins}

\usepackage{color}
\usepackage{ifthen}
\newboolean{editorial}
\setboolean{editorial}{true}
\newcommand{\editorial}[2]{\ifthenelse{\boolean{editorial}}{\textcolor{red}{[\textsf{\textbf{{#1}}}: }\textcolor{blue}{\textsf{{#2}}}\textcolor{red}{]}}{}}

\begin{document}

\title{Departures from the FLRW Cosmological Model in an Inhomogeneous Universe: A Numerical Examination}

\author{John T. Giblin, Jr${}^{1,2}$}
\author{James B. Mertens${}^{2}$}
\author{Glenn D. Starkman${}^{2}$}

\affiliation{${}^1$Department of Physics, Kenyon College, 201 N College Rd, Gambier, OH 43022}
\affiliation{${}^2$CERCA/ISO, Department of Physics, Case Western Reserve University, 10900 Euclid Avenue, Cleveland, OH 44106}

\begin{abstract}
While the use of numerical general relativity for modeling astrophysical phenomena and compact objects is commonplace, the application to cosmological scenarios is only just beginning. Here, we examine the expansion of a spacetime using the Baumgarte-Shapiro-Shibata-Nakamura (BSSN) formalism of numerical relativity in synchronous gauge. This work represents the first numerical cosmological study that is fully relativistic, non-linear and without symmetry.  The universe that emerges exhibits an average Friedmann-Lema\"itre-Robertston-Walker (FLRW) behavior, however this universe also exhibits locally inhomogeneous expansion beyond that expected in linear perturbation theory around a FLRW background.
\end{abstract}

\maketitle

In the past two decades, numerical general relativity (GR) has been widely applied to astrophysical compact objects. Simulations of neutron stars and black holes \cite{Pretorius:2005gq,Campanelli:2005dd,Baker:2005vv,Garfinkle:2003bb} and, very recently, scalar fields \cite{Clough:2015sqa}, have provided answers to old questions about gravity. The success of the BSSN formalism in stabilizing error growth in numerical GR has made it a standard by which numerical GR results are measured \cite{Babiuc:2007vr}. Further, the ability to perform fully non-linear simulations of GR have allowed us to better understand weak gravity, understand where linearized gravity is sufficient and where it breaks down.

On the other hand, current cosmological work typically relies on either a perturbative approach (e.g. \cite{Adamek:2013wja,Adamek:2015eda}) or a Newtonian gravity approximation.
Such work has provided highly precise simulations and resolved how non-linear structure emerges. 
These simulations---as in almost all cosmology---rely on a Friedmann-Lema\"itre-Robertston-Walker (FLRW) cosmological background.  It is commonly assumed that any sub-horizon inhomogeneous structure of the Universe will contribute to an average expansion of the Universe on horizon-sized volumes driven by the horizon-averaged density. When photons are propagated through a simulated universe, they are red-shifted according to the homogeneous FLRW expansion and corrections from Sachs-Wolfe effects.

Such simplified assumptions have long been a matter of concern, and have often been questioned (e.g. \cite{Paranjape:2009zu,Buchert:2011sx}). There have been attempts to address these assumptions
(eg. \cite{Clifton:2009jw,Clifton:2009bp,Clifton:2011mt,Bentivegna:2012ei,Bolejko:2012ue,Clifton:2013jpa}), although for practical reasons such work is typically in an idealized or simplified context. 
The object of this investigation is to begin to move that evaluation, and the quantification of the consequent inaccuracy or imprecision of our cosmological inferences, into a fully general-relativistic setting. For example, most current models decouple the local evolution of matter from the expansion of the universe due to the vast difference of scales, even though the non-linear nature of GR allows power to move between these scales. Determining if a scale exists on which the expansion of the universe can be considered truly homogeneous, despite local variations in curvature, is an open question.

The BSSN formalism \cite{Baumgarte:1998te,Shibata:1995we} is a modification of the ADM Hamiltonian formalism \cite{Arnowitt:1959ah} of GR designed to improve the numerical stability of the latter by introducing auxiliary variables.
The equations that define the BSSN formalism are nonlinear, and therefore formidable to work with analytically. Nevertheless, the nonlinear terms are few enough that -- depending on gauge choice -- numerically integrating the fully unconstrained Einstein equations does not require significantly more computing resources than working in a linearized gravity regime.

In this letter, we study a space-time in which inhomogeneities are present at a range of scales. We set FLRW-like initial conditions -- that is, we include density inhomogeneities with a range of power-spectrum amplitudes on top of a slice of the FLRW metric of constant extrinsic curvature.  We then evolve the complete BSSN dynamical system in the full non-linear GR framework without assuming a background solution.  As the simulation progresses, we monitor the consistency of the usual FLRW approximation by: (1) observing how well the evolution of the fields corresponds to linearized theory; and (2) observing how well the average expansion rate corresponds to expectations for the background.  To our knowledge this is the first simulation of its kind, i.e. the first cosmological work that is fully non-linear, fully relativistic and does not impose symmetries or dimensional-reductions.

Our goal is this: to introduce a methodology that allows us to model fully non-linear gravity on cosmological scales.  We can reproduce homogeneous models of the expanding universe, but also see departures from homogeneity, demonstrating that this method will lead to understanding the future role and need of full GR simulations for cosmological observables.

The software we have developed to simulate cosmological scenarios in full numerical relativity, {\sc CosmoGRaPH}, has passed a standard set of tests and is able to evolve more scenarios than those presented here. For more information about these tests, the full implementation of our numerical method, and future plans for using and for releasing the code, please see our companion paper \cite{Mertens:2015ttp}. In this letter we will focus on the main result of these simulations: the first numerical demonstration of an inhomogeneous, but nearly FLRW, matter-dominated cosmological space time in full GR.

{\sl The BSSN Formalism:} The BSSN formalism parameterizes the spacetime metric by
\begin{equation}
g_{\mu\nu}=\left(\begin{array}{cc}
-\alpha^{2}+\gamma_{lk}\beta^{l}\beta^{k} & \beta_{i}\\
\beta_{j} & \gamma_{ij}\,,
\end{array}\right)
\end{equation}
where we generally refer to $\alpha$ as the `lapse', and $\beta_{i}$ as the `shift'. The metric is rescaled by a conformal factor, $\phi$, so that $\gamma_{ij} = e^{4\phi}\bar{\gamma}_{ij}$, with $\det(\bar{\gamma}_{ij}) = 1$. The components of the 3-metric are then dynamically evolved for a particular gauge choice, along with the extrinsic curvature $K_{ij}$. To do this latter part, the extrinsic curvature is decomposed into a trace part, $K$, and a conformally related trace-free part, $\bar{A}_{ij}$, via $K_{ij} = e^{4\phi}\bar{A}_{ij} + \frac{1}{3}\gamma_{ij}K$, whose indices are raised and lowered by the conformal metric.

The content of the Universe is decomposed into,
\begin{eqnarray}
\rho & = & n_\mu n_\nu T^{\mu\nu}\\
S_i & = & -\gamma_{i\mu} n_\nu T^{\mu\nu} \\
S_{ij} & = & \gamma_{i\mu} \gamma_{j\nu} T^{\mu\nu} \,,
\end{eqnarray}
where $T^{\mu\nu}$ is the stress-energy tensor, $n_\mu = (-\alpha, \vec{0})$ and $S = \gamma^{ij}S_{ij}$.  (For a full textbook treatment, see eg. \cite{BaumgarteShapiroBook}.)

The dynamical equations of motion for the metric are determined by Einstein's equations; however, for stability, the auxiliary conformal connection variables $\bar{\Gamma}^i$ are evolved simultaneously to eliminate terms with mixed derivatives when calculating the Ricci tensor. While {\sc CosmoGRaPH} allows for arbitrary lapse and shift (see \cite{Mertens:2015ttp}), in this letter we employ synchronous gauge (geodesic slicing). In this gauge the lapse is a fixed constant ($\alpha = 1$), and there is no shift ($\beta^i = 0$).
The system we evolve is
\begin{align}
\partial_{t}\phi & =  -\frac{1}{6} K\\
\partial_{t}\bar{\gamma}_{ij} & =  -2\bar{A}_{ij}\\
\partial_{t} K & =  \bar{A}_{ij}\bar{A}^{ij}+\frac{1}{3}K^2+4\pi(\rho+ S)\\
\partial_{t}\bar{A}_{ij} & =  e^{-4\phi}(R_{ij}-8\pi S_{ij})^{TF}+K\bar{A}_{ij}-2\bar{A}_{il}\bar{A}_{j}^{l}\\
\partial_{t}\bar{\Gamma}^{i} & =  2\bar{\Gamma}_{jk}^{i}\bar{A}^{jk}-\frac{4}{3}\bar{\gamma}^{ij}\partial_{j} K-16\pi\bar{\gamma}^{ij}S_{j}+12\bar{A}^{ij}\partial_{j}\phi\,.
\end{align}

For a flat FLRW solution to Einstein's equations, the BSSN variables can be directly translated to FLRW variables. The spatial metric is $\gamma_{ij}^{FLRW} = a^{2}\delta_{ij}$, meaning $\gamma^{FLRW} \equiv \det\gamma_{ij}^{FLRW}=a^{6}$. This relationship, along with our gauge choice, gives us a translation between BSSN and FLRW parameters: $H = -\frac{1}{3}K_{FLRW}$ and $a = \gamma_{FLRW}^{1/6} = e^{2\phi_{FLRW}}$. 

As a proxy for a universe containing matter, we source the metric with a flux-conservative form of the relativistic hydrodynamic equations \cite{Font:2000pp}. In this letter, we restrict ourselves to a $w=0$ cosmological fluid with rest-mass density $\rho_0$ with no initial velocity component;
the contribution to the source terms are then $\rho = \rho_0$, $S_i = 0$ and $S_{ij} = 0$
(and so also $S = 0$).
 The equation of motion for the matter fluid in the absence of initial velocity and for our gauge choice is a simple conservation law,
\begin{equation}
\partial_t \tilde{D} = \partial_{t}(\gamma^{1/2}\rho_0) = 0\,.
\end{equation}

We evolve a finite-volume 3-torus universe with periodic boundary conditions. We set the total volume of the simulation such that the length of any side, $L$, is an arbitrary fraction, $n$, of the initial cosmological horizon in an exact FLRW solution, $L = n H_I^{-1}$. Working in units of $H_I$ fixes the initial energy density of the corresponding FLRW solution to be
\begin{equation}
\rho_{FLRW} = \frac{3}{8\pi}\left(\frac{n}{N \Delta x}\right)^2,
\end{equation}
with $N^3$ the total number of grid points in the volume and $\Delta x = L/N$ the coordinate distance between points. In all of the simulations presented in this letter, we take $n=1/2$, and our units of momentum to be $\Delta k = 2\pi/L$.

We want to be careful about the assumptions of our toy model. First, we utilize periodic boundary conditions as the best choice to reproduce a statistically homogeneous universe on large scales. Such boundary conditions are common in cosmological simulations as they introduce no numerical effects at the boundaries (which can contaminate physics in the light cone of such boundaries).  Since our peak power is at scales much smaller than the length of the side of the box, we do not expect structure to form at wavelengths where periodic effects are significant.  For the time being we also chose two simplifying assumptions about the matter source: its power spectrum and its equation of state.  While these are approximations, they should be sufficiently close to our Universe to yield a reasonable first result (the majority of the energy density of the universe that contributes to the generation of structure is collisionless, pressureless matter).  The initial power spectrum, while only an approximation, is sufficient to show that non-linear effects are relevant.

{\sl Initial Conditions:} The initial surface from which we evolve should satisfy the Hamiltonian and momentum constraint equations
\begin{equation}
\mathcal{H} \equiv
\bar{\gamma}^{ij}\bar{D}_i\bar{D}_j e^\phi - \frac{e^{\phi}}{8}\bar{R} + \frac{e^{5\phi}}{8}\tilde{A}_{ij}\tilde{A}^{ij} - \frac{e^{5\phi}}{12}K^2 + 2\pi e^{5\phi}\rho = 0
\label{hamconst}
\end{equation}
and
\begin{equation}
\mathcal{M}^i = \bar{D}_j (e^{6\phi}\tilde{A}^{ij}) - \frac{2}{3}e^{6\phi}\bar{D}^i K - 8\pi e^{10\phi}S^i = 0\,.
\label{momconst}
\end{equation}
For an FLRW solution, the Hamiltonian constraint is one of the Friedmann equations, and all terms in the momentum constraint equation are zero.

Solving Eqs.~\ref{hamconst} and \ref{momconst} for an arbitrary matter source does not uniquely specify a spatial metric; unconstrained degrees of freedom remain. Making choices that simplify the constraint equations, such as using the conformal transverse-traceless decomposition, can make finding an initial surface easier. Rather than attempting to set initial conditions that perfectly mimic our universe, we opt to obtain a simple solution to Eq.~\ref{hamconst} that is approximately a power-law in momentum space at large and small wavevector $k$, and peaks at a desired scale.

We first specify an extrinsic curvature (akin to the Hubble parameter) approximately determined by the average density, corresponding to an FLRW background. We then introduce fluctuations in the matter field and conformal factor, approximately setting the matter density power spectrum up to that conformal factor.  At large scales (small $k$) the matter power spectrum we choose scales as $P_k \propto k$, and at small scales (large $k$) as $P_k\propto k^{-3}$ \cite{Tegmark:2003ud}. Given a peak scale $k_*$ and corresponding peak amplitude $P_*$, the conformally related matter power spectrum (and approximate matter power spectrum) is then 
\begin{equation}
\label{matterpower}
P_k^{\rho} = \frac{4P_*}{3} \frac{ k/k_*}{1+\frac{1}{3}(k/k_*)^{4}}\,.
\end{equation}
Again, note that this is not intended to perfectly represent our universe, and issues of gauge and conformal rescalings have not been addressed. Also significantly, we introduce a cutoff $k_{\rm cutoff}$ in order to reduce fluctuations on scales where grid effects become important, so that fluctuations are resolved by sufficiently many points. The spectrum is cut using a sigmoid,
\begin{equation}
\label{matterpowercut}
P_k^{\rm cutoff} = \frac{1}{1 + \exp\left[10(k-k_{\rm cutoff})\right]}P_k .
\end{equation}
In this work we take $k_{\rm cutoff}$ to be $10 \Delta k$ so that on a $N^3 = 128^3$ grid we resolve the shortest wavelengths with $12$ grid points for our initial conditions. We find excessive constraint violation for larger cutoffs \cite{Mertens:2015ttp}.

We construct the initial metric by decomposing $\rho$ into $\rho_K$, which sources the trace of the extrinsic curvature $K$, and  $\rho_\psi$, which sources the conformal factor $\psi\equiv e^\phi$. The total density is $\rho = \rho_K + \rho_\psi$. We use a conformally flat metric ($\bar{\gamma}_{ij} = \delta_{ij}$) and set the trace-free part of the extrinsic curvature to zero, leaving us with two simpler equations to solve: 
\begin{eqnarray}
\label{psiic}
\nabla^{2}\psi & = & -2\pi\psi^{5}\rho_{\psi}\\
K & = & -\sqrt{24\pi\rho_{K}}\,.
\end{eqnarray}
Here we choose $K$ to be constant on the initial slice. The  equation for $\psi$ is then difficult to solve for a fixed matter source, with attempted relaxation and iterative solution methods tending to find the $\psi=0$ solution. We therefore create a Gaussian-random realization of the field $\psi$ with a power spectrum $P_k^{\psi} = P_k^{\rho}/k^4$, and then solve for $\rho$.  Note that $\rho_K$ is not necessarily the average density, as $\rho_\psi$ can have nonzero average (although the average of $\psi^{5}\rho_{\psi}$ must be zero for periodic boundary conditions).

In more conceptual terms, we parameterize the spatial distortion of the metric at each point in terms of two parameters. $\phi$ holds information about the volume at that point, $\gamma^{1/2} = e^{6\phi}$; $K$ encodes the rate at which that volume is changing, $\gamma^{-1/2} d\gamma^{1/2}/d t = 3 K$. For a given distribution of matter $\rho$, one has physical (not just gauge) freedom to chose a specific solution for the (initial) values of $\phi$ and $K$.
In the FLRW limit, $\phi$ increases monotonically, and $\phi-\phi_{\mathrm initial}$ corresponds to half the number of elapsed e-foldings, $a_{\rm FLRW} = e^{2\phi_{\rm FLRW}}$.
We therefore use the spatial average value, $\bar{\phi}$, as a proxy for time in most of our plots. This is certainly a good choice if we are close to the FLRW solution.  We define the average to be taken on the constant program-time hypersurfaces (defined as $t={\rm constant}$).  These are not physically important hypersurfaces (see \cite{Buchert:2001sa,Larena:2009md} for discussions on the effects of choosing hypersurfaces when defining averages,); however, they do approach the standard FLRW constant-time hypersurfaces in a homogeneous universe.  

We have examined the accuracy of our model by looking at how well we satisfy the two main constraint equations. Numerically, we employ a diffusive term \cite{Duez:2004uh} to reduce the amount of constraint violation, although it has no significant effect on the quantities of interest in the short term. Details can be found in \cite{Mertens:2015ttp}.

\begin{figure}[hbt]
  \centering
    \includegraphics[width=0.45\textwidth]{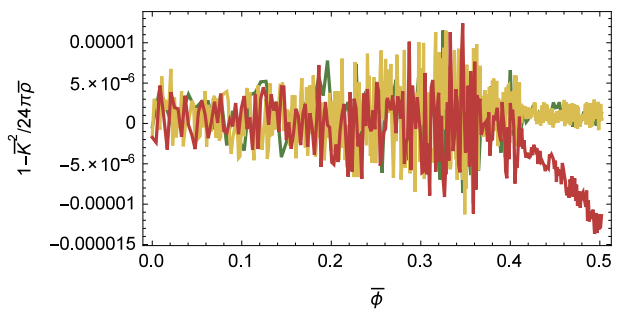}
  \caption{\label{hubbleagreement}
    A comparison of the average value of $K$ for three different resolutions ($64^3$ red, $128^3$ green, and $256^3$ yellow) versus the average value of the conformal factor, $\bar{\phi}$. This simulation has inhomogeneities with $\sigma_{\rho}/\bar{\rho} = 0.05$. }
\end{figure}

{\sl Results:} Having set initial conditions describing a universe expanding at a constant rate across a set of points, each representing slightly different volumes, we can address how well FLRW quantities are recovered in our analysis. To test this we compare the average value of $K$ and $\rho$. In an exact FLRW universe, 
$K = -\sqrt{24\pi\rho}$. We have chosen a large amplitude of inhomogeneity $\sigma_{\rho} / \rho$, but one low enough that no point has  $\rho<0$ in the initial conditions.  In the code we set a value of $P_*$ -- which maps directly to the variation of the density parameter, $\sigma_{\rho}/\bar{\rho}$.  Fig.~\ref{hubbleagreement} shows that for $\sigma_{\rho}/\bar{\rho}= 0.05$, we see excellent agreement with FLRW cosmology for averaged quantities.

\begin{figure*}[htb]
  \centering
    \includegraphics[width=0.45\textwidth]{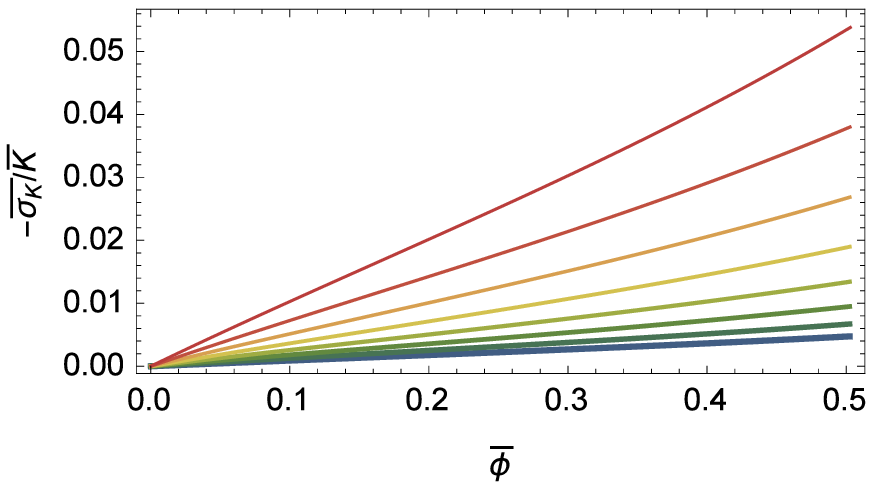}
    \includegraphics[width=0.45\textwidth]{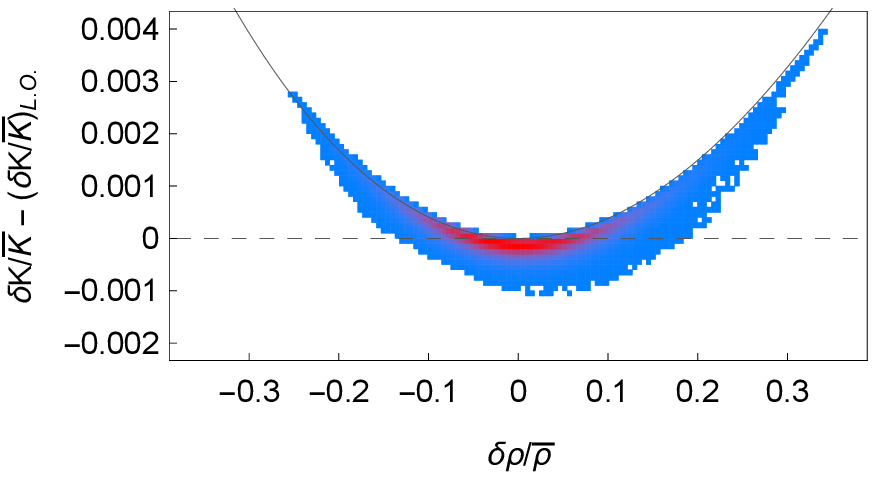}
  \caption{\label{statistics_plot}
The left panel shows variations in the extrinsic curvature, $\sigma_K/\bar{K}$, versus variations in $\sigma_{\rho}/\bar{\rho}$ over the course of a number of runs. The initial $\sigma_{\rho} / \bar{\rho}$ for these runs were $\sigma_{\rho} / \bar{\rho} = 0.009$, $0.0133$, $0.019$, $0.027$, $0.038$, $0.053$, $0.076$, and $0.107$ from bottom to top, or blue to red.
The right panel shows the relationship between fluctuations in matter density and extrinsic curvature at $\bar{\phi} = 0.5$ (one e-fold) as differences from a linear approximation. The dashed line represents the linear-order analytic solution and the solid line shows the solution excluding $\tilde{A}_{ij}\tilde{A}^{ij}$ contributions.  We superimpose a histogram showing all points from a simulation corresponding to an intermediate value of inhomogeneity, $\sigma_\rho/\bar{\rho} = 0.038$, where blue bins contain relatively few points and red bins contain many points. We see local violations of the linear-order approximation of $\mathcal{O}(5\%)$.}
\end{figure*}

If inhomogeneities are important in our volume, we expect that varying $\sigma_{\rho}/\bar{\rho}$ should induce increasingly important variations in $K$.
Fig.~\ref{statistics_plot} shows the reaction of the simulation to values of $\sigma_{\rho}/\bar{\rho}$ between  $\sim2\%$ and  $10\%$.  We see that $\sigma_K/\bar{K}$ grows over the course of the simulations.

Although introducing the numerical method and demonstrating its robustness is the main result of this paper, we note that our simulations go beyond a linear approximation; 
even for relatively small variance in the matter field, $\mathcal{O}(5\%)$ deviations from linear expectations emerge.  This is clear in fig.~\ref{statistics_plot} (right panel) which shows how the local response to the metric is related to local over(under)densities. While we have not yet demonstrated exactly how and if our simulations differ from  fully non-linear Newtonian simulations, we have produced a result that goes beyond perturbative relativistic analyses. 

We can attach some intuition to this result.  We can analytically predict the metric to linear order in perturbations $\delta K$ and $\delta \rho$ around average quantities $\bar{\rho}$ and $\bar{K}$, by writing down the evolution equations
\begin{eqnarray}
\partial_t \delta \rho & = & \bar{\rho} \delta K + \bar{K} \delta \rho \\
\partial_t \delta  K & = & \frac{2}{3} \bar{K} \delta K + 4 \pi \delta \rho \,.
\end{eqnarray}
These are simple ODEs and remind us of those in standard cosmological perturbation theory in Synchronous gauge---see e.g. \cite{Ma:1995ey} where the authors use $\dot{h}$ as the extrinsic curvature, $\delta K = -\dot{h}/2$, and $\delta$ for the density contrast, $\delta = \delta\rho / \rho$. Indeed, so long as any contribution from $\tilde{A}_{ij}\tilde{A}^{ij}$ is negligible, even the full evolution equations for $\rho$ and $K$ remain a set of ODEs that can be integrated easily. The dashed lines in the right panel of Fig.~\ref{statistics_plot} show the behavior of solutions to these equations for our initial conditions. The curves show that the response of the metric is centered about the linear-order predicted value, but exhibit noticeable deviations from it.

\begin{figure}[htb]
  \centering
    \includegraphics[width=0.45\textwidth]{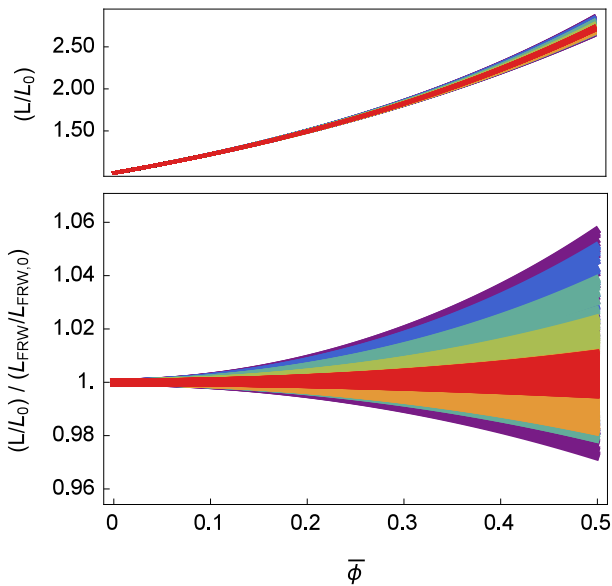}
  \caption{\label{FLRWtest}
The top panel shows ratio of the proper length to the initial length of six sets of $2^7$ paths in our simulation.  The color coding goes from paths of initial coordinate distance of $4\Delta x$ (purple) to paths of initial coordinate distance $128\Delta x$ (red) in spectral order.  The bottom panel shows the same set paths, but compares the ratio of proper length to initial proper length with the FLRW estimate.  We note that shorter paths deviate more from the FLRW expectation.}
\end{figure}

While we have so far demonstrated that we generate variations of the extrinsic curvature, $K$, from point to point, we would like to conclude by performing an additional test.  One of the main predictions of an FLRW universe is that any path on a constant $t$ surface (no matter the shape) has a proper length that scales with the scale factor of the Universe.  Here, we examine whether the expansion of the universe deviates from this expectation.  We will define a set of arbitrary paths on our constant $t$ hyper-surfaces.  If we calculate the proper length of these paths and track the ratio of these lengths as a function of time, we can tell whether we are truly seeing deviations from FLRW behavior. Fig.~\ref{FLRWtest} shows that the growth of the proper length of these paths depends on the length of the path (and not just some scale factor, $a$).  Further, Fig.~\ref{FLRWtest} suggests that this departure is more important at smaller distances than at larger distances.

There is a remaining question of gauge associated with Fig.~\ref{FLRWtest}, as gauge-dependent metric fluctuations can mimic a departure from a homogeneous source. Although we do not compute gauge-independent quantities, the density fluctuations we compute in synchronous gauge will nevertheless source quantities such as optical scalars. Thus, examining path lengths in our spacetimes can provide some intuition into the behavior of observable quantities (details of which we leave for a future study). Here we simply draw a connection to intuition, which suggests that distances of paths with lengths comparable to the Hubble scale should scale as they would in FLRW due to the expansion of the universe being driven by average quantities. If this were true, the longest paths in Fig.~\ref{FLRWtest}, which are approximately half a Hubble length, $H^{-1}/2$, would not diverge from the FLRW expectation. We indeed see longer paths more closely following FLRW behavior; however we also see a departure from FLRW behaviors for individual paths.

{\sl Discussion:} We have seen that the conformal factor varies on sub-horizon scales. These local variations indicate that non-linear gravitational effects are present on cosmological scales. While this should not be surprising, this is the first quantified study of the expansion of an inhomogeneous matter-dominated spacetime within a full, unconstrained General Relativistic framework.
Our companion paper \cite{Mertens:2015ttp} provides further details on our numerical methods and the tests to which we have subjected our code, {\sc CosmoGRaPH}. In future work we expect to report on improved code performance near FLRW solutions, and to quantify the effects on standard physical observables, such as  how photons respond to non-linear gravitational effects in an inhomogeneous universe.

{\sl Acknowledgments}
We thank Eugene Lim, Katy Clough, and Thomas Baumgarte for valuable conversations.  JTG is supported by the National Science Foundation, PHY-1414479; JBM and GDS are supported by a Department of Energy grant DE-SC0009946 to CWRU. We acknowledge the National Science Foundation, the Research Corporation for Science Advancement and the Kenyon College Department of Physics for providing the hardware used to carry out some of these simulations. This work also made use of the High Performance Computing Resource in the Core Facility for Advanced Research Computing at Case Western Reserve University.

\bibliography{references}

\begin{thebibliography}{27}
\expandafter\ifx\csname natexlab\endcsname\relax\def\natexlab#1{#1}\fi
\expandafter\ifx\csname bibnamefont\endcsname\relax
  \def\bibnamefont#1{#1}\fi
\expandafter\ifx\csname bibfnamefont\endcsname\relax
  \def\bibfnamefont#1{#1}\fi
\expandafter\ifx\csname citenamefont\endcsname\relax
  \def\citenamefont#1{#1}\fi
\expandafter\ifx\csname url\endcsname\relax
  \def\url#1{\texttt{#1}}\fi
\expandafter\ifx\csname urlprefix\endcsname\relax\def\urlprefix{URL }\fi
\providecommand{\bibinfo}[2]{#2}
\providecommand{\eprint}[2][]{\url{#2}}

\bibitem[{\citenamefont{Pretorius}(2005)}]{Pretorius:2005gq}
\bibinfo{author}{\bibfnamefont{F.}~\bibnamefont{Pretorius}},
  \bibinfo{journal}{Phys. Rev. Lett.} \textbf{\bibinfo{volume}{95}},
  \bibinfo{pages}{121101} (\bibinfo{year}{2005}), \eprint{gr-qc/0507014}.

\bibitem[{\citenamefont{Campanelli et~al.}(2006)\citenamefont{Campanelli,
  Lousto, Marronetti, and Zlochower}}]{Campanelli:2005dd}
\bibinfo{author}{\bibfnamefont{M.}~\bibnamefont{Campanelli}},
  \bibinfo{author}{\bibfnamefont{C.~O.} \bibnamefont{Lousto}},
  \bibinfo{author}{\bibfnamefont{P.}~\bibnamefont{Marronetti}},
  \bibnamefont{and}
  \bibinfo{author}{\bibfnamefont{Y.}~\bibnamefont{Zlochower}},
  \bibinfo{journal}{Phys. Rev. Lett.} \textbf{\bibinfo{volume}{96}},
  \bibinfo{pages}{111101} (\bibinfo{year}{2006}), \eprint{gr-qc/0511048}.

\bibitem[{\citenamefont{Baker et~al.}(2006)\citenamefont{Baker, Centrella,
  Choi, Koppitz, and van Meter}}]{Baker:2005vv}
\bibinfo{author}{\bibfnamefont{J.~G.} \bibnamefont{Baker}},
  \bibinfo{author}{\bibfnamefont{J.}~\bibnamefont{Centrella}},
  \bibinfo{author}{\bibfnamefont{D.-I.} \bibnamefont{Choi}},
  \bibinfo{author}{\bibfnamefont{M.}~\bibnamefont{Koppitz}}, \bibnamefont{and}
  \bibinfo{author}{\bibfnamefont{J.}~\bibnamefont{van Meter}},
  \bibinfo{journal}{Phys. Rev. Lett.} \textbf{\bibinfo{volume}{96}},
  \bibinfo{pages}{111102} (\bibinfo{year}{2006}), \eprint{gr-qc/0511103}.

\bibitem[{\citenamefont{Garfinkle}(2004)}]{Garfinkle:2003bb}
\bibinfo{author}{\bibfnamefont{D.}~\bibnamefont{Garfinkle}},
  \bibinfo{journal}{Phys. Rev. Lett.} \textbf{\bibinfo{volume}{93}},
  \bibinfo{pages}{161101} (\bibinfo{year}{2004}), \eprint{gr-qc/0312117}.

\bibitem[{\citenamefont{Clough et~al.}(2015)\citenamefont{Clough, Figueras,
  Finkel, Kunesch, Lim, and Tunyasuvunakool}}]{Clough:2015sqa}
\bibinfo{author}{\bibfnamefont{K.}~\bibnamefont{Clough}},
  \bibinfo{author}{\bibfnamefont{P.}~\bibnamefont{Figueras}},
  \bibinfo{author}{\bibfnamefont{H.}~\bibnamefont{Finkel}},
  \bibinfo{author}{\bibfnamefont{M.}~\bibnamefont{Kunesch}},
  \bibinfo{author}{\bibfnamefont{E.~A.} \bibnamefont{Lim}}, \bibnamefont{and}
  \bibinfo{author}{\bibfnamefont{S.}~\bibnamefont{Tunyasuvunakool}}
  (\bibinfo{year}{2015}), \eprint{1503.03436}.

\bibitem[{\citenamefont{Babiuc et~al.}(2008)}]{Babiuc:2007vr}
\bibinfo{author}{\bibfnamefont{M.~C.} \bibnamefont{Babiuc}}
  \bibnamefont{et~al.}, \bibinfo{journal}{Class. Quant. Grav.}
  \textbf{\bibinfo{volume}{25}}, \bibinfo{pages}{125012}
  (\bibinfo{year}{2008}), \eprint{0709.3559}.

\bibitem[{\citenamefont{Adamek et~al.}(2013)\citenamefont{Adamek, Daverio,
  Durrer, and Kunz}}]{Adamek:2013wja}
\bibinfo{author}{\bibfnamefont{J.}~\bibnamefont{Adamek}},
  \bibinfo{author}{\bibfnamefont{D.}~\bibnamefont{Daverio}},
  \bibinfo{author}{\bibfnamefont{R.}~\bibnamefont{Durrer}}, \bibnamefont{and}
  \bibinfo{author}{\bibfnamefont{M.}~\bibnamefont{Kunz}},
  \bibinfo{journal}{Phys. Rev.} \textbf{\bibinfo{volume}{D88}},
  \bibinfo{pages}{103527} (\bibinfo{year}{2013}), \eprint{1308.6524}.

\bibitem[{\citenamefont{Adamek et~al.}(2015)\citenamefont{Adamek, Daverio,
  Durrer, and Kunz}}]{Adamek:2015eda}
\bibinfo{author}{\bibfnamefont{J.}~\bibnamefont{Adamek}},
  \bibinfo{author}{\bibfnamefont{D.}~\bibnamefont{Daverio}},
  \bibinfo{author}{\bibfnamefont{R.}~\bibnamefont{Durrer}}, \bibnamefont{and}
  \bibinfo{author}{\bibfnamefont{M.}~\bibnamefont{Kunz}}
  (\bibinfo{year}{2015}), \eprint{1509.01699}.

\bibitem[{\citenamefont{Paranjape}(2009)}]{Paranjape:2009zu}
\bibinfo{author}{\bibfnamefont{A.}~\bibnamefont{Paranjape}}
  (\bibinfo{year}{2009}), \eprint{0906.3165}.

\bibitem[{\citenamefont{Buchert and Räsänen}(2012)}]{Buchert:2011sx}
\bibinfo{author}{\bibfnamefont{T.}~\bibnamefont{Buchert}} \bibnamefont{and}
  \bibinfo{author}{\bibfnamefont{S.}~\bibnamefont{Räsänen}},
  \bibinfo{journal}{Ann. Rev. Nucl. Part. Sci.} \textbf{\bibinfo{volume}{62}},
  \bibinfo{pages}{57} (\bibinfo{year}{2012}), \eprint{1112.5335}.

\bibitem[{\citenamefont{Clifton and
  Ferreira}(2009{\natexlab{a}})}]{Clifton:2009jw}
\bibinfo{author}{\bibfnamefont{T.}~\bibnamefont{Clifton}} \bibnamefont{and}
  \bibinfo{author}{\bibfnamefont{P.~G.} \bibnamefont{Ferreira}},
  \bibinfo{journal}{Phys. Rev.} \textbf{\bibinfo{volume}{D80}},
  \bibinfo{pages}{103503} (\bibinfo{year}{2009}{\natexlab{a}}),
  \bibinfo{note}{[Erratum: Phys. Rev.D84,109902(2011)]}, \eprint{0907.4109}.

\bibitem[{\citenamefont{Clifton and
  Ferreira}(2009{\natexlab{b}})}]{Clifton:2009bp}
\bibinfo{author}{\bibfnamefont{T.}~\bibnamefont{Clifton}} \bibnamefont{and}
  \bibinfo{author}{\bibfnamefont{P.~G.} \bibnamefont{Ferreira}},
  \bibinfo{journal}{JCAP} \textbf{\bibinfo{volume}{0910}}, \bibinfo{pages}{026}
  (\bibinfo{year}{2009}{\natexlab{b}}), \eprint{0908.4488}.

\bibitem[{\citenamefont{Clifton et~al.}(2012)\citenamefont{Clifton, Ferreira,
  and O'Donnell}}]{Clifton:2011mt}
\bibinfo{author}{\bibfnamefont{T.}~\bibnamefont{Clifton}},
  \bibinfo{author}{\bibfnamefont{P.~G.} \bibnamefont{Ferreira}},
  \bibnamefont{and}
  \bibinfo{author}{\bibfnamefont{K.}~\bibnamefont{O'Donnell}},
  \bibinfo{journal}{Phys. Rev.} \textbf{\bibinfo{volume}{D85}},
  \bibinfo{pages}{023502} (\bibinfo{year}{2012}), \eprint{1110.3191}.

\bibitem[{\citenamefont{Bentivegna and Korzynski}(2012)}]{Bentivegna:2012ei}
\bibinfo{author}{\bibfnamefont{E.}~\bibnamefont{Bentivegna}} \bibnamefont{and}
  \bibinfo{author}{\bibfnamefont{M.}~\bibnamefont{Korzynski}},
  \bibinfo{journal}{Class. Quant. Grav.} \textbf{\bibinfo{volume}{29}},
  \bibinfo{pages}{165007} (\bibinfo{year}{2012}), \eprint{1204.3568}.

\bibitem[{\citenamefont{Bolejko and Ferreira}(2012)}]{Bolejko:2012ue}
\bibinfo{author}{\bibfnamefont{K.}~\bibnamefont{Bolejko}} \bibnamefont{and}
  \bibinfo{author}{\bibfnamefont{P.~G.} \bibnamefont{Ferreira}},
  \bibinfo{journal}{JCAP} \textbf{\bibinfo{volume}{1205}}, \bibinfo{pages}{003}
  (\bibinfo{year}{2012}), \eprint{1204.0909}.

\bibitem[{\citenamefont{Clifton et~al.}(2013)\citenamefont{Clifton, Gregoris,
  Rosquist, and Tavakol}}]{Clifton:2013jpa}
\bibinfo{author}{\bibfnamefont{T.}~\bibnamefont{Clifton}},
  \bibinfo{author}{\bibfnamefont{D.}~\bibnamefont{Gregoris}},
  \bibinfo{author}{\bibfnamefont{K.}~\bibnamefont{Rosquist}}, \bibnamefont{and}
  \bibinfo{author}{\bibfnamefont{R.}~\bibnamefont{Tavakol}},
  \bibinfo{journal}{JCAP} \textbf{\bibinfo{volume}{1311}}, \bibinfo{pages}{010}
  (\bibinfo{year}{2013}), \eprint{1309.2876}.

\bibitem[{\citenamefont{Baumgarte and Shapiro}(1998)}]{Baumgarte:1998te}
\bibinfo{author}{\bibfnamefont{T.~W.} \bibnamefont{Baumgarte}}
  \bibnamefont{and} \bibinfo{author}{\bibfnamefont{S.~L.}
  \bibnamefont{Shapiro}}, \bibinfo{journal}{Phys. Rev.}
  \textbf{\bibinfo{volume}{D59}}, \bibinfo{pages}{024007}
  (\bibinfo{year}{1998}), \eprint{gr-qc/9810065}.

\bibitem[{\citenamefont{Shibata and Nakamura}(1995)}]{Shibata:1995we}
\bibinfo{author}{\bibfnamefont{M.}~\bibnamefont{Shibata}} \bibnamefont{and}
  \bibinfo{author}{\bibfnamefont{T.}~\bibnamefont{Nakamura}},
  \bibinfo{journal}{Phys. Rev.} \textbf{\bibinfo{volume}{D52}},
  \bibinfo{pages}{5428} (\bibinfo{year}{1995}).

\bibitem[{\citenamefont{Arnowitt et~al.}(1959)\citenamefont{Arnowitt, Deser,
  and Misner}}]{Arnowitt:1959ah}
\bibinfo{author}{\bibfnamefont{R.~L.} \bibnamefont{Arnowitt}},
  \bibinfo{author}{\bibfnamefont{S.}~\bibnamefont{Deser}}, \bibnamefont{and}
  \bibinfo{author}{\bibfnamefont{C.~W.} \bibnamefont{Misner}},
  \bibinfo{journal}{Phys. Rev.} \textbf{\bibinfo{volume}{116}},
  \bibinfo{pages}{1322} (\bibinfo{year}{1959}).

\bibitem[{\citenamefont{Mertens et~al.}(2015)\citenamefont{Mertens, Giblin, and
  Starkman}}]{Mertens:2015ttp}
\bibinfo{author}{\bibfnamefont{J.~B.} \bibnamefont{Mertens}},
  \bibinfo{author}{\bibfnamefont{J.~T.} \bibnamefont{Giblin}},
  \bibnamefont{and} \bibinfo{author}{\bibfnamefont{G.~D.}
  \bibnamefont{Starkman}} (\bibinfo{year}{2015}), \eprint{1511.01106}.

\bibitem[{\citenamefont{Baumgarte and Shapiro}(2010)}]{BaumgarteShapiroBook}
\bibinfo{author}{\bibfnamefont{T.~W.} \bibnamefont{Baumgarte}}
  \bibnamefont{and} \bibinfo{author}{\bibfnamefont{S.~L.}
  \bibnamefont{Shapiro}}, \emph{\bibinfo{title}{Numerical Relativity: Solving
  Einstein's Equations on the Computer}} (\bibinfo{publisher}{Cambridge
  University Press}, \bibinfo{address}{Cambridge, UK}, \bibinfo{year}{2010}).

\bibitem[{\citenamefont{Font}(2000)}]{Font:2000pp}
\bibinfo{author}{\bibfnamefont{J.~A.} \bibnamefont{Font}},
  \bibinfo{journal}{Living Rev. Rel.} \textbf{\bibinfo{volume}{3}},
  \bibinfo{pages}{2} (\bibinfo{year}{2000}), \bibinfo{note}{[Living Rev.
  Rel.6,4(2003)]}, \eprint{gr-qc/0003101}.

\bibitem[{\citenamefont{Tegmark et~al.}(2004)}]{Tegmark:2003ud}
\bibinfo{author}{\bibfnamefont{M.}~\bibnamefont{Tegmark}} \bibnamefont{et~al.}
  (\bibinfo{collaboration}{SDSS}), \bibinfo{journal}{Phys. Rev.}
  \textbf{\bibinfo{volume}{D69}}, \bibinfo{pages}{103501}
  (\bibinfo{year}{2004}), \eprint{astro-ph/0310723}.

\bibitem[{\citenamefont{Buchert}(2001)}]{Buchert:2001sa}
\bibinfo{author}{\bibfnamefont{T.}~\bibnamefont{Buchert}},
  \bibinfo{journal}{Gen. Rel. Grav.} \textbf{\bibinfo{volume}{33}},
  \bibinfo{pages}{1381} (\bibinfo{year}{2001}), \eprint{gr-qc/0102049}.

\bibitem[{\citenamefont{Larena}(2009)}]{Larena:2009md}
\bibinfo{author}{\bibfnamefont{J.}~\bibnamefont{Larena}},
  \bibinfo{journal}{Phys. Rev.} \textbf{\bibinfo{volume}{D79}},
  \bibinfo{pages}{084006} (\bibinfo{year}{2009}), \eprint{0902.3159}.

\bibitem[{\citenamefont{Duez et~al.}(2004)\citenamefont{Duez, Shapiro, and
  Yo}}]{Duez:2004uh}
\bibinfo{author}{\bibfnamefont{M.~D.} \bibnamefont{Duez}},
  \bibinfo{author}{\bibfnamefont{S.~L.} \bibnamefont{Shapiro}},
  \bibnamefont{and} \bibinfo{author}{\bibfnamefont{H.-J.} \bibnamefont{Yo}},
  \bibinfo{journal}{Phys. Rev.} \textbf{\bibinfo{volume}{D69}},
  \bibinfo{pages}{104016} (\bibinfo{year}{2004}), \eprint{gr-qc/0401076}.

\bibitem[{\citenamefont{Ma and Bertschinger}(1995)}]{Ma:1995ey}
\bibinfo{author}{\bibfnamefont{C.-P.} \bibnamefont{Ma}} \bibnamefont{and}
  \bibinfo{author}{\bibfnamefont{E.}~\bibnamefont{Bertschinger}},
  \bibinfo{journal}{Astrophys. J.} \textbf{\bibinfo{volume}{455}},
  \bibinfo{pages}{7} (\bibinfo{year}{1995}), \eprint{astro-ph/9506072}.

\end{thebibliography}
 
\end{document}